\documentclass[aps,prc,twocolumn,showpacs,superscriptaddress,floatfix]{revtex4-1}
\usepackage{graphicx}
\usepackage{bm}
\usepackage{amssymb}
\usepackage{hyperref}
\usepackage{color}

\hypersetup{
	colorlinks=true,
	pdfborder={0,0,0}.
	citecolor=blue,
	linkcolor=blue,
	urlcolor=blue
	}
	
\newcommand{\CTwoN}{$^{12}$C($^6$He,$^4$He)$^{14}$C}

\begin{document}
\author{D.~Smalley}\thanks{Present Address: The National Superconducting Cyclotron Laboratory, Michigan State University, East Lansing, MI 48824, US}\email{smalleyd@nscl.msu.edu}\affiliation{Department of Physics, Colorado School of Mines, Golden, CO 80401, USA}
\author{F.~Sarazin}\affiliation{Department of Physics, Colorado School of Mines, Golden, CO 80401, USA}
\author{F.M.~Nunes}\affiliation{National Superconducting Cyclotron Laboratory, Michigan State University, East Lansing, MI 48824, USA}\affiliation{Department of Physics and Astronomy, Michigan State University, East Lansing, MI 48824, USA}
\author{B.A.~Brown}\affiliation{National Superconducting Cyclotron Laboratory, Michigan State University, East Lansing, MI 48824, USA}\affiliation{Department of Physics and Astronomy, Michigan State University, East Lansing, MI 48824, USA}
\author{P.~Adsley}\affiliation{Department of Physics, University of York, York YO10 5DD, United Kingdom}
\author{H.~Al-Falou}\affiliation{TRIUMF, Vancouver V6T 2A3, BC, Canada} 
\author{C.~Andreoiu}\affiliation{Department of Chemistry, Simon Fraser University, Burnaby V5A1S6, BC, Canada}
\author{B.~Baartman}\affiliation{Department of Chemistry, Simon Fraser University, Burnaby V5A1S6, BC, Canada}
\author{G.C.~Ball}\affiliation{TRIUMF, Vancouver V6T 2A3, BC, Canada}
\author{J.C.~Blackmon}\affiliation{Department of Physics and Astronomy, Louisiana State University, Baton Rouge, LA 70803, USA}
\author{H.C.~Boston}\affiliation{Department of Physics, University of Liverpool, Liverpool L69 3BX, United Kingdom}
\author{W.N.~Catford}\affiliation{Department of Physics, University of Surrey, Guildford, GU15XH, United Kingdom} 
\author{S.~Chagnon-Lessard}\affiliation{Department of Physics, University of Guelph, Guelph N1G 2W1, ON, Canada}
\author{A.~Chester}\affiliation{Department of Chemistry, Simon Fraser University, Burnaby V5A1S6, BC, Canada}
\author{R.M.~Churchman}\affiliation{TRIUMF, Vancouver V6T 2A3, BC, Canada}
\author{D.S.~Cross}\affiliation{Department of Chemistry, Simon Fraser University, Burnaby V5A1S6, BC, Canada}
\author{C.Aa.~Diget}\affiliation{Department of Physics, University of York, York YO10 5DD, United Kingdom}
\author{D.~Di Valentino}\affiliation{Department of Physics, Carleton University, Ottawa K1S 5B6, ON, Canada}
\author{S.P.~Fox}\affiliation{Department of Physics, University of York, York YO10 5DD, United Kingdom} 
\author{B.R.~Fulton}\affiliation{Department of Physics, University of York, York YO10 5DD, United Kingdom}
\author{A.~Garnsworthy}\affiliation{TRIUMF, Vancouver V6T 2A3, BC, Canada}
\author{G.~Hackman}\affiliation{TRIUMF, Vancouver V6T 2A3, BC, Canada}
\author{U.~Hager}\affiliation{Department of Physics, Colorado School of Mines, Golden, CO 80401, USA}
\author{R.~Kshetri}\affiliation{Department of Chemistry, Simon Fraser University, Burnaby V5A1S6, BC, Canada}
\author{J.N.~Orce}\affiliation{TRIUMF, Vancouver V6T 2A3, BC, Canada}
\author{N.A.~Orr}\affiliation{LPC-ENSICAEN, IN2P3-CNRS et Universit\'e de Caen, 14050 Caen cedex, France}
\author{E.~Paul}\affiliation{Department of Physics, University of Liverpool, Liverpool L69 3BX, United Kingdom}
\author{M.~Pearson}\affiliation{TRIUMF, Vancouver V6T 2A3, BC, Canada}
\author{E.T.~Rand}\affiliation{Department of Physics, University of Guelph, Guelph N1G 2W1, ON, Canada}
\author{J.~Rees}\affiliation{Department of Physics, University of Liverpool, Liverpool L69 3BX, United Kingdom}
\author{S.~Sjue}\affiliation{TRIUMF, Vancouver V6T 2A3, BC, Canada}
\author{C.E.~Svensson}\affiliation{Department of Physics, University of Guelph, Guelph N1G 2W1, ON, Canada}
\author{E.~Tardiff}\affiliation{TRIUMF, Vancouver V6T 2A3, BC, Canada}
\author{A. Diaz Varela}\affiliation{Department of Physics, University of Guelph, Guelph N1G 2W1, ON, Canada}
\author{S.J.~Williams}\affiliation{TRIUMF, Vancouver V6T 2A3, BC, Canada}
\author{S.~Yates}\affiliation{Department of Physics, University of Kentucky, Lexington, KY 40506, USA}

 \title{Two-neutron transfer reaction mechanisms in $^{12}$C($^6$He,$^{4}$He)$^{14}$C using a realistic three-body $^{6}$He model}
 \date{\today}
             
\begin{abstract}

The reaction mechanisms of the two-neutron transfer reaction $^{12}$C($^6$He,$^4$He) have been studied at 30 MeV at the TRIUMF ISAC-II facility using the SHARC charged-particle detector array.  Optical potential parameters have been extracted from the analysis of the elastic scattering angular distribution.  The new potential has been applied to the study of the transfer angular distribution to the 2$^+_2$ 8.32 MeV state in $^{14}$C, using a realistic 3-body $^6$He model and advanced shell model calculations for the carbon structure, allowing to calculate the relative contributions of the simultaneous and sequential two-neutron transfer.  The reaction model provides a good description of the 30 MeV data set and shows that the simultaneous process is the dominant transfer mechanism.  Sensitivity tests of optical potential parameters show that the final results can be considerably affected by the choice of optical potentials.  A reanalysis of data measured previously at 18 MeV however, is not as well described by the same reaction model, suggesting that one needs to include higher order effects in the reaction mechanism.\\

\end{abstract}

\maketitle
\section{Introduction}
\label{intro}
One of the critical ingredients to understand nuclear properties, both in the valley of stability and at the nuclear drip lines, is the pairing effect \cite{review-pairing}. 
Pairing is a general term that embodies the correlation between pairs of nucleons, producing for example the well-known mass staggering in nuclear isobars. 
Pairing is also essential to understanding the formation of two-neutron halos~\cite{Tan13}.  
Although the importance of pairing for describing nuclear phenomena is well accepted, reaction probes used to measure pairing are still poorly understood.

Two-nucleon transfer is the traditional probe to study pairing.  The main idea is that the angular distribution for the simultaneous transfer of two nucleons depends directly on the change in angular momentum of the two nucleons from the original to the final nucleus, and therefore provides indirectly information on their relative motion. 
Experimental studies of two-neutron transfer imply the use of a surrogate reaction.
Hence, the interpretation of the results become strongly dependent on the reaction mechanism, since typically the simultaneous transfer of the two nucleons is contaminated by the two-step sequential transfer.  

Traditionally A$(t,p)$B or B$(p,t)$A reactions have been the most common tools to explore two-neutron correlations (see for example \cite{decowski,bjerregaard} for earlier studies  and \cite{wimmer,potel11} for more recent studies).  The advantage of $(t,p)$ is that it allows the study of not only the ground state of B but also of a number of its excited states, accessible through energy and angular momentum matching.  Since the triton is a well understood nucleus, it was thought that these reactions would be easier to describe than others using heavier probes \cite{Gat75,Iga91}.  However, missing factors of 2 or 3 in the cross section normalization (known as the unhappiness factor \cite{Mer79, Kun82}) for $(p,t)$ and $(t,p)$ have shown that a simple perturbative description, that does not take into account the intermediate deuteron state correctly has severe limitations \cite{Tho12}.  
There are experimental drawbacks as well: (p,t) only permits the study of the ground-state of the original nucleus considered, and $(t,p)$ requires handling of tritium radioactivity, which is challenging in most laboratories.

Other two-neutron transfer probes have been considered. The next simplest case after $(t,p)$ would be the ($^{6}$He,$^{4}$He) reaction involving the two-neutron halo nucleus $^{6}$He. The structure of $^{6}$He makes it a very attractive candidate for two-neutron transfer reactions, because of its Borromean nature and its very low two-neutron separation energy ($S_{2n}=0.97$ MeV). Indeed, Chatterjee {\it et al.} \cite{Cha08} recently observed in $^{6}$He on $^{65}$Cu at 23~MeV a large dominance of the two-neutron over one-neutron transfer cross-section and interpreted this in terms of the unique features of the $^{6}$He wavefunction. At present, $^6$He is the best understood two-neutron halo nucleus, with a very significant  component where the halo neutrons are spatially correlated (the so-called "di-neutron" component) and an equally important component where the two halo neutrons are anti-correlated (the so-called "cigar" component) (e.g. \cite{Var94,Bri10}).
Given the comparatively small $^6$He two-neutron separation energy with respect to the one of the triton ($S_{2n}=6.25$ MeV),  the two-neutron transfer reaction ($^{6}$He,$^{4}$He) provides not only a higher Q-value overall than its $(t,p)$ counterpart allowing for higher excited states to be populated, but also a more favorable Q-value matching condition for a given two-neutron transfer reaction.  It has even been suggested by Fortunato {\it et al.}~\cite{For02} that this higher Q-value provides relatively large cross sections to study giant pairing vibrations in heavy-ions.  It also provides a different angular momentum matching condition, given that the two active neutrons can exist in a relative p-state as opposed to the situation in the triton. In addition to these two important differences, the Borromean nature of $^6$He \cite{Zhu93} implies that the sequential transfer can only happen through the continuum states of $^5$He and is likely to leave a softer imprint in the distributions than the sequential process in $(t,p)$ through the deuteron bound state. 

Experimentally, the major drawback lies in the fact that $^{6}$He is unstable (T$_{1/2}$=807 ms) and therefore the reaction can only be studied using targets made of stable or long-lived isotopes. Theoretically, one difficulty arises from the existence of unbound excited states in $^{6}$He, the first one (2$^{+}$) at 1.8 MeV.  With this in mind, what is needed to explore ($^{6}$He,$^{4}$He) is a test-case, where the final state is well understood, such that the focus can be on the reaction mechanism.

Given the interest in the halo structure of $^6$He \cite{Tan85,Zhu93}, there have been many measurements  involving the ($^{6}$He,$^{4}$He) vertex. The measurements of the p($^{6}$He,$^{4}$He)t at 151 MeV \cite{wolski-he6pt151,giot-he6pt150}  resulted in difficulties in the analysis due to the strong interference of the small $t+t$ component in the $^6$He wave function. There were also two measurements on $^4$He($^{6}$He,$^{4}$He)$^6$He, one at E$_{\mathrm{lab}}=29$ MeV \cite{raabe-he6he4-11} and the other at E$_{\mathrm{lab}}=151$ MeV \cite{terakopian-he6he4-151}.  In this reaction, the transfer channel corresponds to the exchange of the elastic channel, reducing the number of effective interactions necessary to describe the process, but introducing yet another complication, that of appropriate symmetrization. And while Ref~\cite{Cha08} clearly demonstrated a preference for two-neutron over one-neutron transfer, the actual mechanism for the two-neutron transfer was simply assumed to be one-step arising from the di-neutron configuration of the $^{6}$He wavefunction.

Our test-case is the reaction $^{12}$C($^{6}$He,$^{4}$He)$^{14}$C, a reaction that populates well-known states in $^{14}$C.  Possible intermediate states in $^{13}$C in the sequential transfer are also well known. If we thus assume that our present knowledge of $^6$He is complete, we can focus on the reaction mechanism. $^{12}$C($^{6}$He,$^{4}$He)$^{14}$C was first performed at E=5.9 MeV \cite{Ost98} and populated the ground state and the first $1^-$ and $3^-$ states in $^{14}$C. Following that measurement, the reaction was remeasured at 18 MeV \cite{Mil04}, populating strongly the $8.32$ MeV $2^+_2$ state in $^{14}$C. At the lowest beam energy, the transfer cannot be treated perturbatively, resulting in an intricate mechanism for the process \cite{Kro00},  in \cite{Mil04} a reasonable description of the reaction is provided with the simple one-step di-neutron model. 

In this work, we present  new results from the measurement of $^{12}$C($^{6}$He,$^{4}$He)$^{14}$C at $30$ MeV and discuss in detail the reaction mechanisms taking into account not only one-step but also two-step (through intermediate $^{13}$C states) processes. The new model is also applied to the previous study at 18~MeV. In Section \ref{setup}, we describe the experimental setup and details in the analysis. In Section \ref{theo}, we briefly summarize the main ingredients used in the reaction theory used to analyze our results. The elastic scattering results are presented in Section \ref{elas}, followed by the inelastic scattering in Section \ref{Inelas} and the transfer results in Section \ref{tran}. Finally, in Section \ref{concl} we summarize and draw our conclusions.

\section{Experimental Setup}
\label{setup}

The \CTwoN{} experiment was performed in direct kinematics at the TRIUMF ISAC-II facility using a combination of the Silicon Highly-segmented Array for Reactions and Coulex (SHARC) \cite{Dig11} and the TRIUMF-ISAC Gamma-Ray Escape-Suppressed Spectrometer (TIGRESS) \cite{Hac13}.  The $^6$He$^{+}$ beam was produced by impinging a 500-MeV, 75-$\mu$A proton beam upon a 20.63~g/cm$^{2}$ ZrC target, and extracted using a Forced Electron Beam Induced Arc Discharge (FEBIAD) ion source. The 12.24~keV $^{6}$He beam was post-accelerated through the ISAC-I (where the beam was also stripped to 2+) and ISAC-II accelerators to 30 MeV before being delivered to the TIGRESS beam line. A small amount ($<$5\%) of $^6$Li contaminant was also transmitted through the accelerator. The beam then impinged upon a 217~$\mu$g/cm$^2$ $^{12}$C target located at the center of the SHARC and TIGRESS arrays. The average beam intensity was estimated to be $\mathrm{I}=8\times10^5$ pps with a total integrated beam current on target of 77~nC over the course of the experiment.  

\begin{figure}
\centering
\includegraphics[width=\columnwidth]{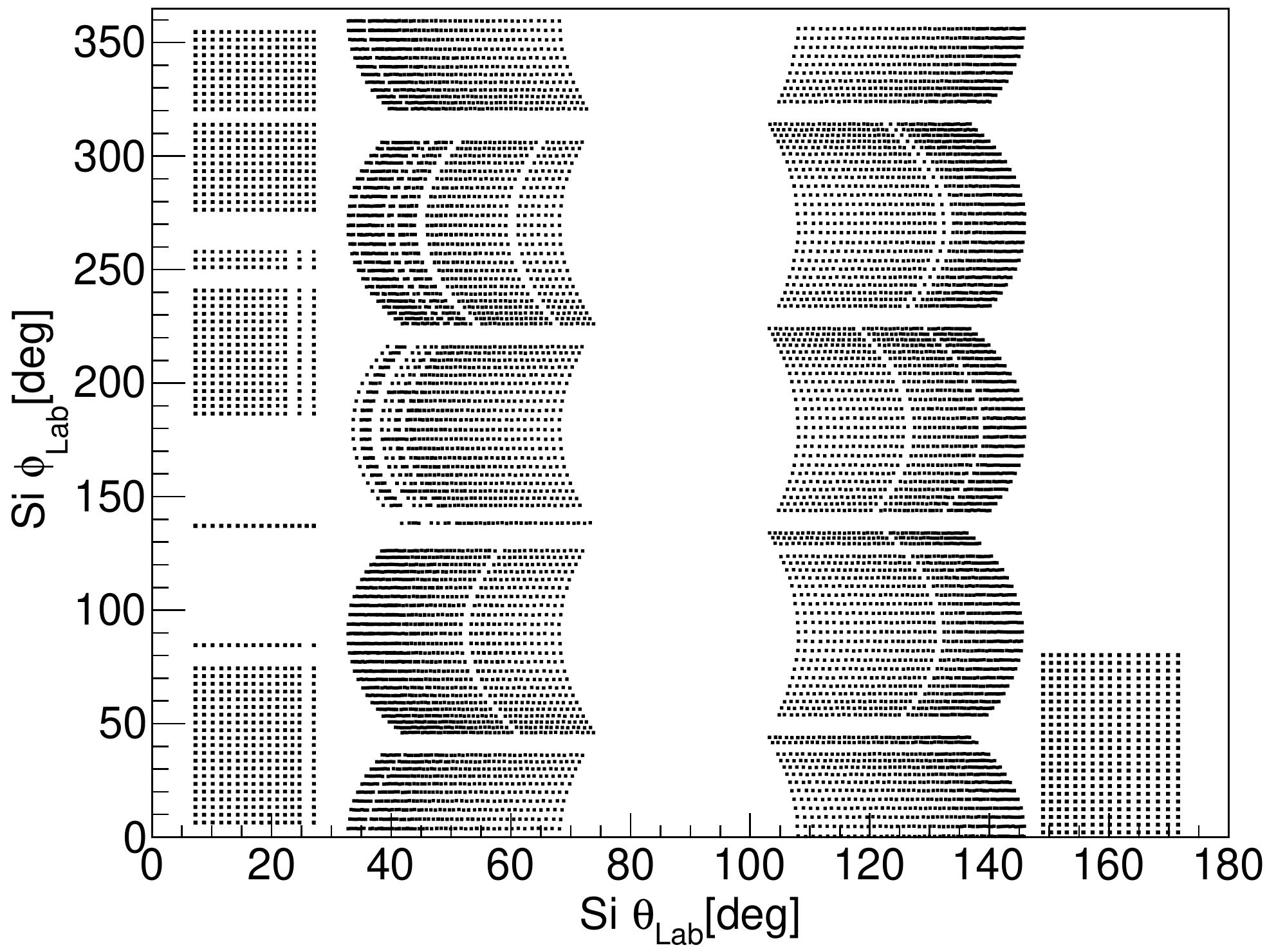}
\caption{\small Laboratory angular coverage of SHARC for this experiment.  The areas of detection are: the downstream end cap (DCD) at $7^\circ<\theta_{\mathrm{lab}}<27^\circ$, the downstream box (DBx) at $32^\circ<\theta_{\mathrm{lab}}<71^\circ$, the upstream box (UBx) at $103^\circ<\theta_{\mathrm{lab}}<145^\circ$ and the partial upstream end cap (UCD) at $148^\circ<\theta_{\mathrm{lab}}<172^\circ$. One of the DCD detector was single-sided, hence the single row of pixels along the $\theta_{lab}$ direction at $135^\circ<\phi_{lab}<140^\circ$ which depicts the center $\phi_{\mathrm{lab}}$ angle used for reconstruction (the coverage was $100^\circ<\phi_{lab}<175^\circ$).}
\label{f:SHARC}

\includegraphics[width=\columnwidth]{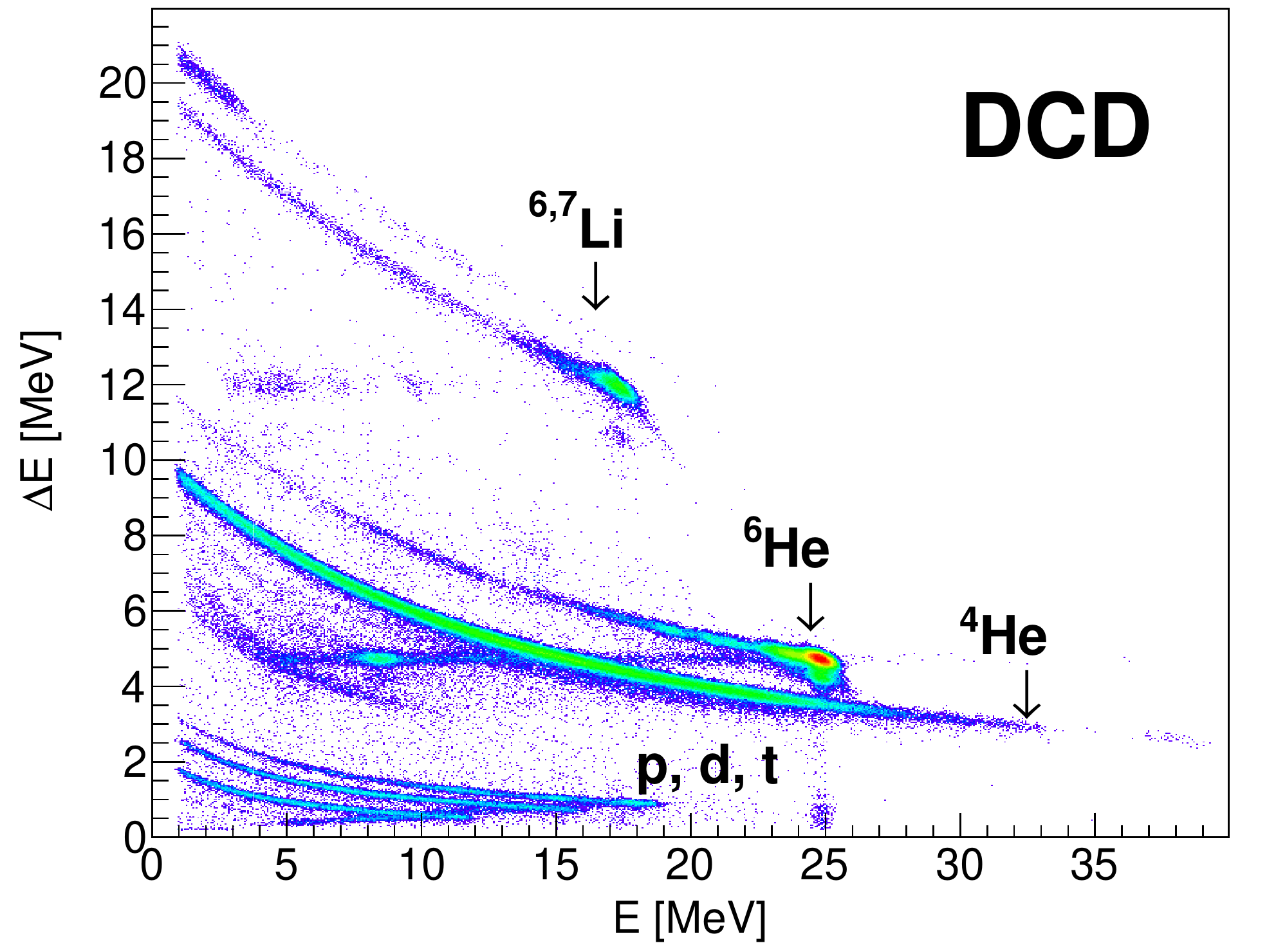}
\caption{\small (color online) Particle identification for the DCD.  The $\Delta$E-E plot shows the energy loss in the thin 80 $\mu$m detector ($\Delta$E) plotted against the energy loss in the thick 1mm detector (E).  The nuclei unambiguously identified are shown in the figure.}
\label{f:ParID}
\end{figure}
The laboratory angular coverage of the SHARC array is shown in Fig.~\ref{f:SHARC}, where each point represents a pixel of detection. The downstream end cap detectors (DCD) consisted of four $\Delta$E-E telescopes made of three 80 $\mu$m Double-Sided Silicon Strip Detectors (DSSSD) and one single-sided 40 $\mu$m $\Delta$E detector, all backed with 1 mm E detectors.  Polar angle coverage in the laboratory frame of the DCD was 7$^\circ$-27$^\circ$.  The downstream box (DBx) was comprised of four 140 $\mu$m DSSSDs $\Delta$E detectors backed with 1.5 mm E detectors and polar angle coverage of 32$^\circ$-71$^\circ$.  The upstream box (UBx) was made of four 1 mm DSSSDs with polar angle coverage of 103$^\circ$-145$^\circ$.  Finally, the partial upstream end cap (UCD) consisted of a single 1 mm DSSSD with a polar angle coverage of 148$^\circ$-172$^\circ$.
A total of 21 strips out of 752 were not functioning giving a solid angle coverage of the array of $\Omega\approx2\pi$. Additionally, one DCD E ($184^\circ<\phi_{lab}<258^\circ$) detector malfunctioned during the experiment.

Particle identification was obtained for three quadrants of DCD and all quadrants of DBx.  The DCD $\Delta$E-E provided identification of all the isotopes of charge $\mathrm{Z}\leq3$ as shown in Fig.~\ref{f:ParID}. In particular, clear separation of the $^4$He and $^6$He isotopes was achieved allowing for a clear identification of the elastic/inelastic scattering from the two-neutron transfer (and fusion-evaporation) channels. Only identification of particles with mass $\mathrm{A}\leq4$ was possible in the downstream box due to the thickness of the $\Delta$E detectors (which stopped the scattered $^6$He beam). For UBx and UCD, identification of the two-neutron transfer channel remained possible because of the high Q-value of the reaction (Q=12.15~MeV).

The TIGRESS array was used in high-efficiency mode for $\gamma$-ray detection. The angular coverage of the TIGRESS array was such that there were 7 clover detectors at $\theta_{\mathrm{lab}}=90^\circ$ and 4 clover detectors at $\theta_{\mathrm{lab}}=135^\circ$ with one crystal not operational in the clover at ($\theta_{lab}=135^\circ$,$\phi_{lab}=113^\circ$). Detection of $\gamma$-rays in coincidence with charged-particles was achieved, but the statistics was too low to carry out analyses of $\gamma$-gated angular distributions. In what follows, $\gamma$-ray detection and tagging are not discussed further.

\section{Theory}
\label{theo}
\subsection{Reaction mechanisms}
We study two-neutron transfer with the finite-range Distorted Wave Born Approximation  (DWBA). The simultaneous two-neutron transfer process is treated to first order, as one step (see e.g. \cite{oganessian99} for details).
The transition amplitude for the process for $^{12}$C($^6$He,$^4$He)$^{14}$C$(2^+)$ can be written as \cite{oganessian99}:
\begin{equation}
T^{post} = \langle \chi_f \: {\cal I}_{^{14}C,^{12}C} | \Delta V |  {\cal I}_{^6He,^4He} \: \chi_i\rangle \;,
\label{t-matrix}
\end{equation} 
where $\chi_i$ and $\chi_f$ are the initial and final distorted waves between $^6$He-$^{12}$C and $^4$He-$^{14}$C respectively, and ${\cal{I}}_{He}$ and ${\cal{I}}_C$ are the two-neutron overlap functions of the ground state of $^6$He and $^4$He, and the $2^+_2$ state of $^{14}$C and the ground state of $^{12}$C respectively. In the post form \cite{Tho09}, the potential is defined as, 
\begin{equation}
\Delta V = V_{2n^{12}C}+V_{\alpha^{12}C}-U_i,
\label{V}
\end{equation} 
where $V_{2n^{12}C}$ is the potential between the two valence neutrons of the $^6$He and the $^{12}$C, $V_{\alpha^{12}C}$ is the core-core potential and $U_i$ is the entrance channel potential ($^6$He+$^{12}$C).

Many analyses of two-neutron transfer data involving $^6$He beams have assumed that the $^6$He system can be described by a two-body wave function of the $\alpha$-particle and a di-neutron cluster, simplifying tremendously the two-neutron overlap function needed in the transfer matrix element (e.g. \cite{Cha08,Mil04}). While a full six-body microscopic description \cite{Bri10,timofeyuk01} may not provide more than a correction to the overall normalization of the overlap function, the three-body $^6\mathrm{He}$~=~$^{4}\mathrm{He}+\mathrm{n}+\mathrm{n}$ wave function is a requirement for a consistent treatment of simultaneous and sequential transfer, present for all but the highest energies E$>$100~MeV. In this work, we use a realistic three-body model for $^{6}$He \cite{Bri08} which reproduces the binding energy and radius of the ground state. The two-neutron spectroscopic amplitudes for $^{14}$C are obtained from microscopic shell-model calculations, which we describe in more detail below. We assume a standard geometry (radius $r=1.25$ fm and diffuseness $a=0.65$ fm) for the mean field generating the radial form factor including a spin-orbit with the same geometry as the mean field and a strengh of V = 6.5 MeV for the $^{14}$C two-neutron overlap function. 

The initial and final distorted waves are also important elements of Eq.~\ref{t-matrix}. One way to constrain these is by using elastic scattering over a wide angular range. For the optical potential in the final channel $^4$He-$^{14}$C, there are many possible data sets from which to draw, and even global parameterizations may be adequate since none of the nuclei involved are of peculiar structure. The same is not true for the initial channel. Separate studies on the elastic scattering of $^6$He on $^{12}$C \cite{Alk96,Lap02}  at high energies have revealed very significant modifications of the expected optical potential based on the double folding approach, due to large breakup effects inherent to the low S$_{2n}$ of $^{6}$He. Given the strong dependence of the two-neutron transfer cross section on the optical potentials used,  it is critical to have elastic scattering data at the appropriate energy for a meaningful analysis. This is presented in Section~\ref{elas}.

At the energies we are interested in, the two-step sequential process must be considered. In this case we need to include the one-nucleon transfer into $^{13}$C:
\begin{equation}
T^{post} = \langle \chi_{f1} \; {\cal I}_{^{13}C,^{12}C} | \Delta V_1 |  {\cal I}_{^6He,^5He}\; \chi_{i1}\rangle \;,
\label{t-matrix1}
\end{equation} 
where the initial state of Eq.~\ref{t-matrix} is taken into a bound state in $^{13}$C and followed by the second neutron transferring from $^5$He to the final state in $^{14}$C:
\begin{equation}
T^{prior} = \langle \chi_{f2}\; {\cal I}_{^{14}C,^{13}C} | \Delta V_2 |  {\cal I}_{^5He,^4He} \;\chi_{i2}\rangle \;,
\label{t-matrix2}
\end{equation} 

A number of $^{13}$C states contribute to this process, and the needed spectroscopic amplitudes are obtained from microscopic shell model calculations, assuming the same residual interaction and model space as that used for the two-neutron amplitudes.  We avoid the explicit inclusion of states in the continuum by modifying the binding energy and level scheme of $^{13}$C. 
\subsection{Shell model considerations for carbon overlaps}
All structural information for the carbon isotopes relies on recent shell model predictions.
The earliest discussion related to the structure of the 2$^+$ 8.32 MeV state in $^{14}$C is based on the gamma decay of its analogue in $^{14}$N at 10.43 MeV \cite{war60}. There are two 2$^+$ T=1 states in this energy region of $^{14}$N, the 10.43 MeV and a lower one at 9.16 MeV.  The possible shell-model configurations for these states relative to a closed shell for $^{16}$O are two-holes in the $p$-shell ($2h$) and four holes in the $p$-shell with two particles in the $sd$ shell ($2p-4h$). The electromagnetic decay required about an equal admixture between these two configurations. The early weak-coupling model of Lie \cite{lie72} could reproduce this result only by introducing an empirical energy shift in the $2p-4h$ component. Mordechai {\it et al.} \cite{mor78} have used the empirical wave functions of Lie to understand the relative $^{12}$C$(t,p)$$^{14}$C strength to the 7.01 and 8.32 MeV 2$^+$ states. The $(t,p)$ cross section is dominated by the $2p$ ($sd$) part of the transfer, and the mixing results in about equal $(t,p)$ cross sections for these two 2$^+$ states. 

Here we are able to use the full $p-sd$ model space. In this model space the basis dimension is 982,390 for $J=2$, and the wavefunctions can be obtained with the NuShellX code \cite{nushellx}.  A Hamiltonian for this space was recently developed by Utsuno and Chiba \cite{uts11} (psdUC).  This was used to calculate Gamow-Teller strengths for the $^{12}$B($^{7}$Li,$^{7}$Be)$^{12}$Be reaction. Relative to the original Hamiltonian of Utsuno and Chiba, the $p-sd$ gap had to be increased by one MeV in order to reproduce the correct mixing of states in $^{12}$Be. We calculated the two-particle transfer amplitudes with this same modified psdUC Hamiltonian. The wavefunctions for the two 2$^+$ states in $^{14}$C came out with about the correct mixing between $2h$ and $2p-4h$ that are required to reproduce the gamma decay and the relative $(t,p)$ cross sections. The theoretical energy splitting of 0.51 MeV is smaller than the experimental value of 1.3 MeV. But the most important aspect for the present analysis is that the structure of the 2$^+$ state be consistent with the history of its structure that we have outlined above.

\subsection{Details of the calculations}
\begin{figure}
\centering
\includegraphics[width=\columnwidth]{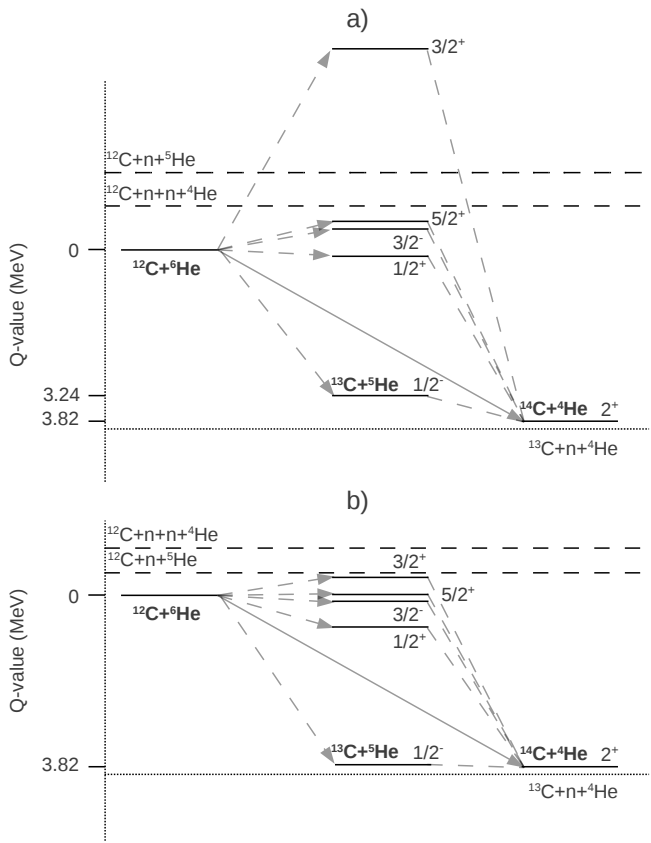}
\caption{\small The level scheme adopted for this two-neutron transfer study.  The solid line shows the pathway for simultaneous transfer. The dotted lines show the multiple pathways taken into account in the sequential transfer.  The total transfer calculation then takes the coherent sum of all the pathways.  The true level scheme is shown in a) and the modified level scheme which was used is depicted in  b). }
\label{f:ModelAssume}
\end{figure}

The level scheme relevant for the transfer is shown in Fig.~\ref{f:ModelAssume} a) with the Q-values listed. 
A direct transfer from the ground state of $^6$He to the $2^{+}_{2}$ 8.32 MeV state in $^{14}$C is shown with a solid line.  The sequential transfer paths, represented by dashed lines, involve single particle overlaps.  To reduce the complexity of the reaction theory and associated computational cost, we  make  a quasi-bound approximation for $^5$He, and  take for the $^5$He optical potentials, the same parameters used for $^6$He-$^{12}$C.  The binding energy of the quasi-bound $^5$He was assumed to be half the binding energy of $^6$He relative to $^4$He (28.7831 MeV). Similarly, we slightly shifted  the intermediate $^{13}$C states, to avoid introducing the continuum in our calculations. The only significant shift introduced was for the high lying $3/2^+$ state. Although the relative spectroscopic amplitudes for the $^{13}$C $3/2^+$ state is weak compared to the most dominant states of 5/2$^+$ and 1/2$^+$, we found that this state has a non-negligible contribution to the cross section.   The total transfer calculation is then the coherent sum of all the pathways.  Fig.~\ref{f:ModelAssume} b) shows the final level scheme adopted in our reaction calculation.  The indirect route of the unbound 2$^+$ excited state of $^6$He was not considered.  Krouglov {\it et al.} \cite{Kro00} observed that in the analysis of the $^6$He+$^{12}$C two-neutron transfer at E$_{\mathrm{lab}}=5.9$ MeV it was of minor importance.

The transfer calculations were performed using the reaction code  {\sc fresco} \cite{Tho88} and the realistic three-body calculations for $^6$He were performed using {\sc efadd} \cite{efadd}.  The integration was performed out to 30 fm and the transfer matrix element non-locality was calculated out to 10 fm.  The cluster variable for the two-neutron overlap extended to 10 fm and partial waves up to J$_{\mathrm{max}}=33$ were included, for convergence.
\section{Results and Discussion}
\subsection{Elastic Scattering}
\label{elas}
\begin{figure}
\centering
\includegraphics[width=\columnwidth]{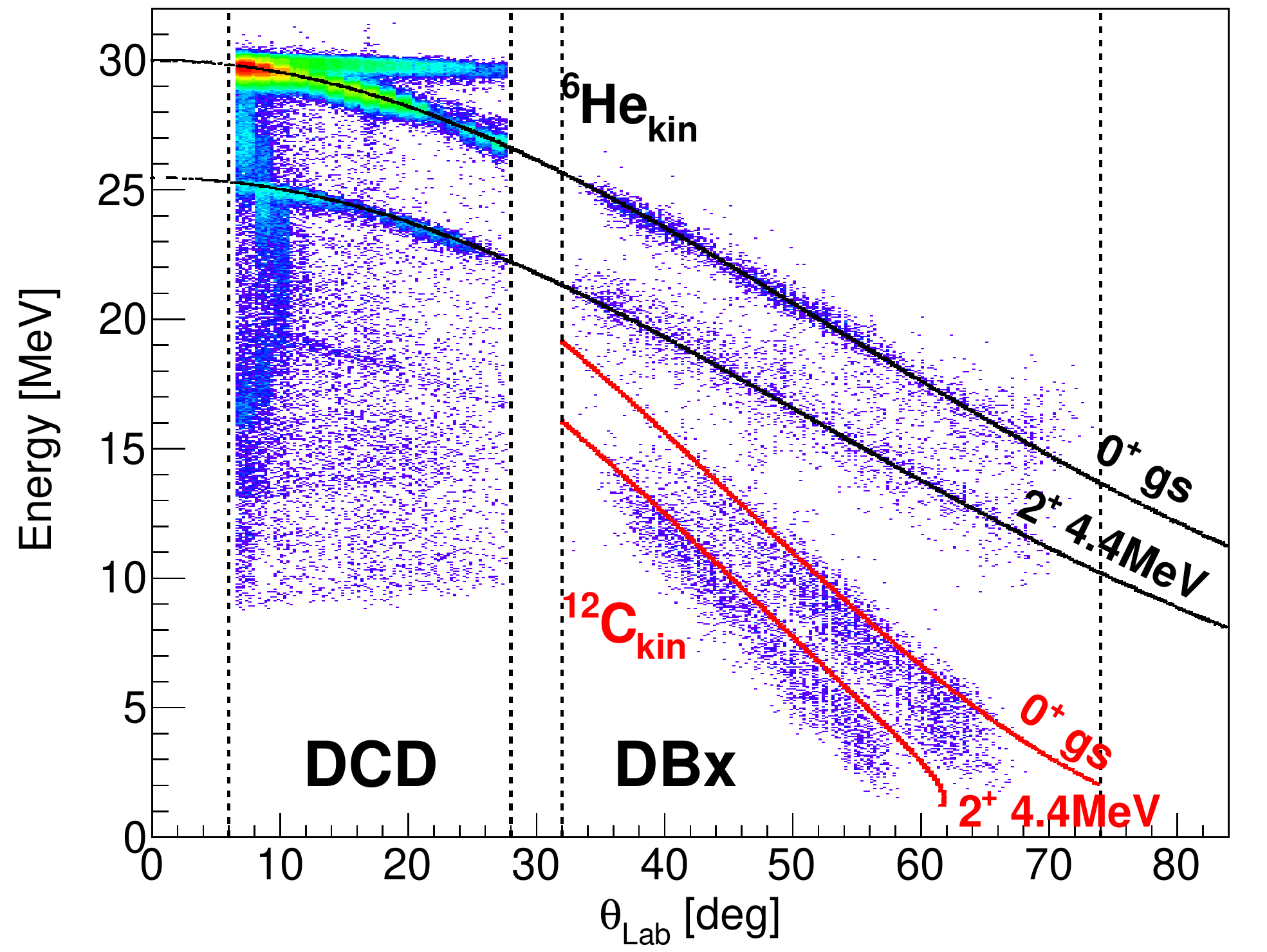}
\caption{\small (color online) The elastic and inelastic scattering Energy vs. $\theta_{\mathrm{lab}}$ kinematics.  The DCD used the straightforward particle ID of the $^6$He ejectiles, while the DBx required coincident detection considerations (see text for details).  The kinematic curves for $^6$He and $^{12}$C are shown in black and red respectively with strong population of the $0^+$ ground state and the $2^+$ 4.4 MeV state of $^{12}$C.  The slight mismatch in energy of the $^{12}$C expected kinematics is due to increased energy loss from the higher Z and straggling in the carbon target.  }
\label{f:Elastic_E_Theta}
\end{figure}
\begin{table*}[t]
\small
\caption{Optical model parameters used in this work.  Potential depths (V, W, V$_{\mathrm{SO}}$) are in units of MeV, radii (r$_{\mathrm{C}}$, r$_{\mathrm{R}}$, r$_{\mathrm{I}}$, r$_{\mathrm{SO}}$) and diffuseness (a$_{\mathrm{R}}$, a$_{\mathrm{I}}$, a$_{\mathrm{SO}}$) are in units of fm.  All referenced radius units are $\mathrm{R}_{\mathrm{x}}=\mathrm{r}_{\mathrm{x}}\mathrm{A}_{\mathrm{t}}^{1/3}$, where A$_{\mathrm{t}}$ corresponds to the number of nucleons in the target (t).}
\label{t:OM-Parameters}
\begin{tabular}{l|*{13}{c}}
\hline
&			r$_{\mathrm{C}}$			&			V			&			r$_{\mathrm{R}}$			&			a$_{\mathrm{R}}$			&			W			&			r$_{\mathrm{I}}$			&			a$_{\mathrm{I}}$			&			V$_{\mathrm{SO}}$			&			r$_{\mathrm{SO}}$			&			a$_{\mathrm{SO}}$\\
\hline
$^6$He+$^{12}$C/$^5$He+$^{13}$C				&            2.3			&                240.3                       &                 1.18                   &              0.74                      &            10.0                       &            2.10                       &            1.18                       &	 				&						&			\\	
$^6$Li+$^{12}$C \cite{Trc90}				&            2.3			&                240.3                       &                 1.18                   &              0.74                      &            10.0                       &            2.10                       &            0.78                       &				&					&			\\
$\alpha$+$^{12,14}$C \cite{Mik61}				&            1.2			&                40.69                       &                 2.12                   &              0.1                      &            2.21                       &            2.98                       &            0.22                       &						&						&			\\	
n+$^{5}$He \cite{Kee08}				&       1.2     			&                4.3			&			1.25			&              0.65                      &                                   &                                  &                                   &			6			&			1.25			&			0.65\\	
\hline
\end{tabular}
\end{table*} 
The elastic scattering angular distribution was extracted for the downstream ($\mathrm{\theta}_{\mathrm{lab}}<90^\circ$) detection system.  For $\mathrm{\theta}_{\mathrm{lab}}>90^\circ$, no clear identification of the elastic scattering channel could be achieved. Fig.~\ref{f:Elastic_E_Theta} shows the Energy vs. $\mathrm{\theta}_{\mathrm{lab}}$ in the DCD and DBx for the $^6$He+$^{12}$C elastic and inelastic scattering after cuts on the data. In the DCD, unambiguous identification of the scattered $^6$He was achieved through $\Delta$E-E particle identification. The elastic angular distribution was only extracted past $\theta_{\mathrm{lab}}$=10.7$^\circ$, because of the presence of elastic scattering on a heavy nucleus in the first three polar strips. As mentioned earlier, (in)elastically scattered $^{6}$He do not punch-through the $\Delta$E in the DBx. However, clean identification of the scattering events could still be achieved using kinematics considerations. For this particular reaction, both the scattered $^{6}$He and recoiling $^{12}$C remain confined to the DBx and are detected in a reaction plane intersecting opposite DBx quadrants. Using cuts on the kinematic correlation of the $^{6}$He and $^{12}$C energies, polar angles $\theta$ and azimuthal angles $\phi$, the kinematic loci for $^{6}$He and $^{12}$C were extracted from background and are shown in Fig.~\ref{f:Elastic_E_Theta}.  The detection of $^{12}$C in the DBx stops abruptly at $\theta_{\mathrm{lab}}=36^\circ$ due to the $^6$He scattering beyond $\theta_{\mathrm{lab}}=72^\circ$. Monte Carlo simulations using GEANT4 \cite{Ago03} were performed to determine the detection efficiency arising from these cuts and to normalize the DCD and DBx relative angular distributions.

\begin{figure}
\centering
\includegraphics[width=\columnwidth]{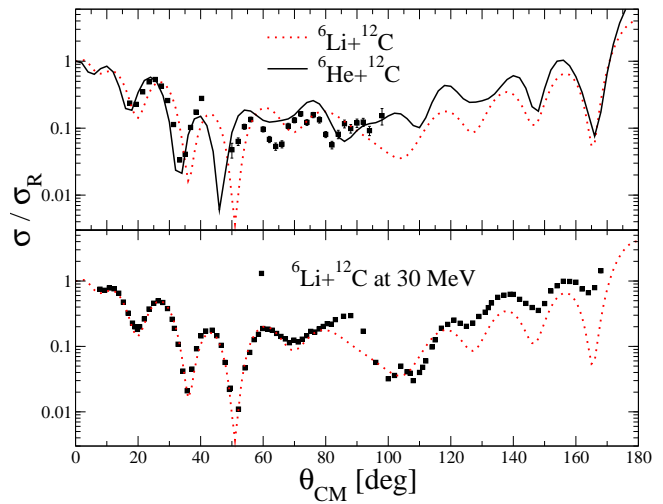}
\caption{\small (color online) The elastic scattering angular distribution.  The top figure shows $^6$He+$^{12}$C elastic scattering at E$_{\mathrm{lab}}=30$ MeV.  The dotted red line is the theoretical cross section calculated using the optical model parameters for the $^6$Li+$^{12}$C at 30 MeV \cite{Trc90}, while the solid line is the theoretical cross section calculated using the optical model parameters of $^6$He+$^{12}$C obtained from the fit to the current data set.  The bottom figure shows the $^6$Li+$^{12}$C elastic scattering at E$_{\mathrm{lab}}=30$ MeV with the dotted red line being the optical model parameters for the $^6$Li+$^{12}$C at 30 MeV \cite{Trc90}.}
\label{f:ElasticAngDist}
\end{figure}

The elastic scattering angular distribution is shown in Fig.~\ref{f:ElasticAngDist} (top).  The absolute cross section was obtained by normalizing the angular distribution to Coulomb scattering at forward angles.  Only statistical errors are shown but we estimate a systematic error in the normalization of 25\%. The solid line is the optical model obtained from a $\chi^2$ minimization of the $^6$Li+$^{12}$C optical model parameters to better fit the current data set, using the code {\sc sfresco}.  We found that adjusting only the imaginary diffuseness $a_I$ of the original $^6$Li potential and introducing the correct Coulomb charge, a good description of the $^6$He was obtained (see solid line in Fig.~\ref{f:ElasticAngDist} (top)).  Similar adjustments were performed by Sakaguchi {\it et al.} \cite{Sak11} to the $^6$Li+p optical potential to describe the $^6$He+p scattering.  We note that, while our $^6$He scattering is measured out to $\theta_{\mathrm{cm}}<100^\circ$, 
the elastic scattering data for  $^6$Li+$^{12}$C  \cite{Vin84} (see Fig.~\ref{f:ElasticAngDist} (bottom))  goes out to much larger angles ($5^\circ<\theta_{\mathrm{cm}}<170^\circ$), and these large angles appear to be critical to better constrain the optical potential.  The optical model parameters used for the elastic scattering are shown in table \ref{t:OM-Parameters}.  

General features of the scattering can be observed from the $^6$Li+$^{12}$C optical model \cite{Trc90} at 30 MeV (see Fig.~\ref{f:ElasticAngDist} (bottom)).  However, the current data has a notable shift in minima compared to the $^6$Li+$^{12}$C optical potential.  This same effect was observed in the elastic scattering at 18 MeV of $^6$He+$^{12}$C by Milin {\it et al.} \cite{Mil04}.  Using the adjusted elastic potential parameters, an analysis was performed for the inelastic scattering to the 2$^{+}$ 4.4~MeV state of $^{12}$C. 
\subsection{Inelastic Scattering (2$^{+}$ 4.4~MeV state)}
\label{Inelas}
\begin{figure}
\centering
\includegraphics[width=\columnwidth]{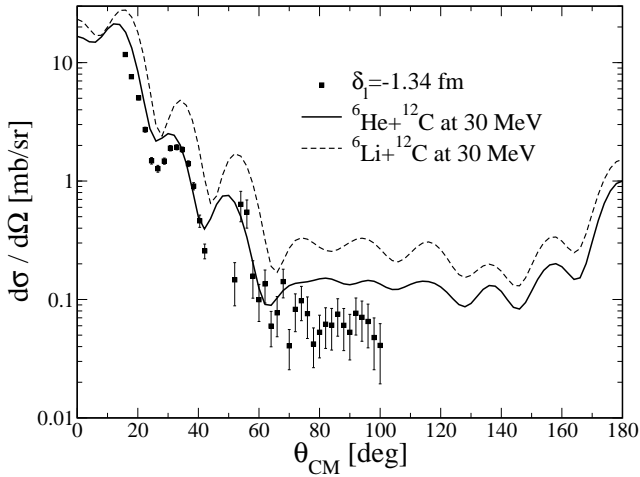}
\caption{\small The inelastic scattering to the 2$^+$ 4.4 MeV state.  An independent model of the transfer was performed with the elastic scattering optical potential described above (see text for details). }
\label{f:inelastic}
\end{figure}

Inelastic scattering of the 2$^+$ 4.4~MeV state of $^{12}$C was observed and is shown in Fig.~\ref{f:Elastic_E_Theta}.  The data were extracted and analyzed independently of the transfer model.  A quadrupole deformation of $^{12}$C was assumed with a deformation length of $\delta_{l}=-1.34$ fm taken from the analysis of inelastic scattering of $^{12}$C+$^6$Li at 30 MeV \cite{Ker96}.  The coupling strength is calculated within the simple rotor model. A coupled-channel calculation with coupling between the ground state and first excited state of $^{12}$C was assumed.  The inelastic angular distribution is shown in Fig.~\ref{f:inelastic}.  The adjustment of the imaginary diffuseness $a_I$ as described in Section \ref{elas}, has the effect of decreasing the magnitude of the angular distribution at higher angles, improving the agreement with the data.   In this work we do not couple the inelastic scattering contributions into the two-neutron transfer.
%
\subsection{${}^{12}$C(${}^6$He,${}^4$He)${}^{14}$C (2$^+$, 8.32 MeV)}
\label{tran}
The two-neutron transfer was extracted beyond background for all areas of detection except the UCD.  Interference with the (p,t) reaction kinematics was observed and this led to the exclusion of events between $10.7^\circ<\theta_{\mathrm{lab}}<19^\circ$.  The DCD/DBx relied on particle identification of the alpha particles from the $\Delta$E-E spectrum.  Coincident events were selected for the DBx in a similar manner to that discussed in Section~\ref{elas}.  Background in the DBx and UBx was present due to a worsening of the angular resolution which was accounted for and subtracted from the spectrum.

\begin{figure}
\centering
\includegraphics[width=\columnwidth]{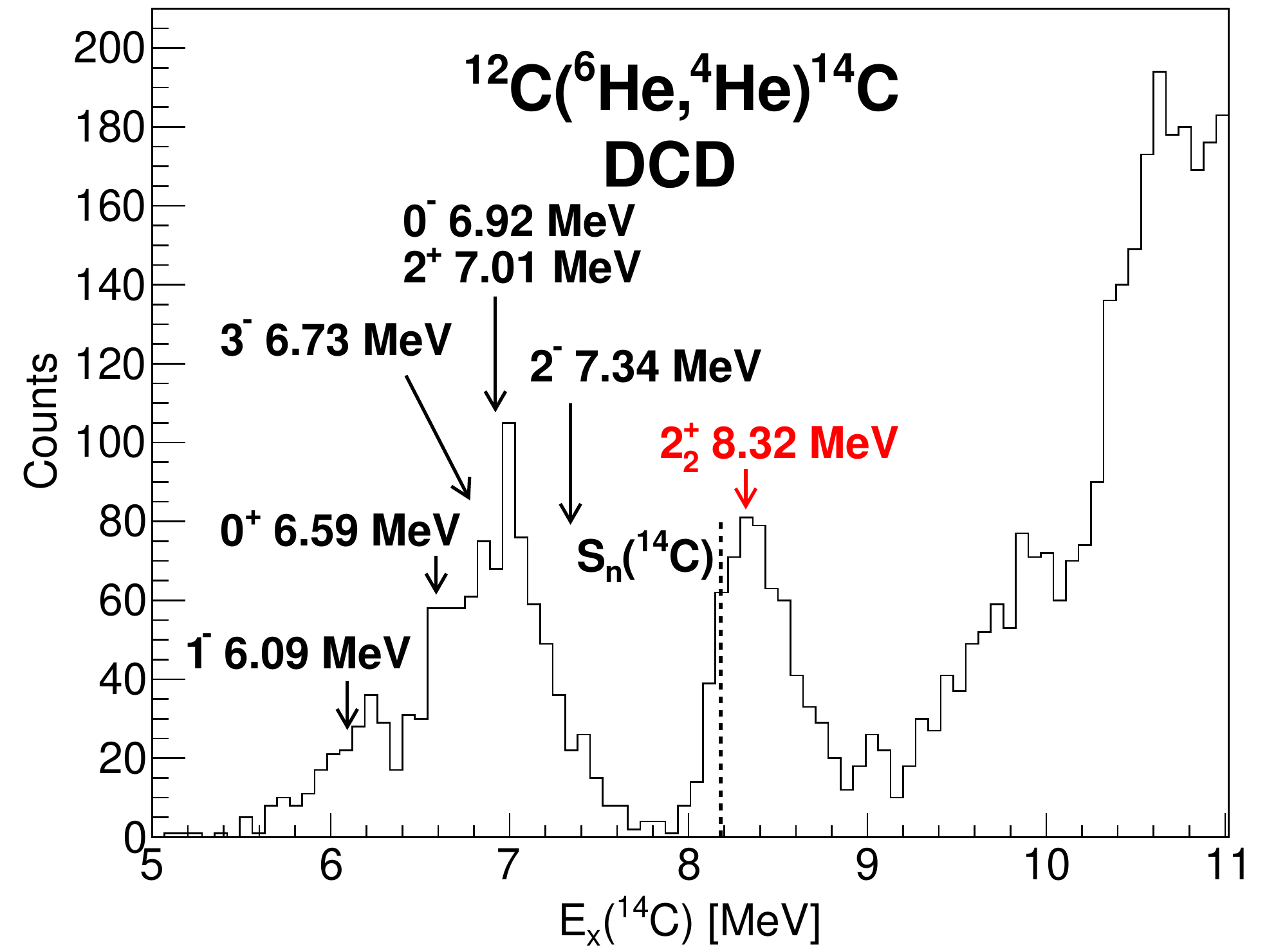}
\caption{\small (color online) The energy excitation spectrum for the two-neutron transfer in the DCD.  Separation of the 2$^+_2$ 8.32 MeV of $^{14}$C is clearly observed.  A high density of bound states from 6 MeV to 7.3 MeV is observed but cannot be separated without $^4$He-$\gamma$ coincidences.  The 8.32 MeV state is located 200 keV above the neutron separation energy ($\mathrm{S}_{n}=8.1$ MeV). }
\label{f:TwoNExcite}
\end{figure}

The excitation spectrum of the two-neutron transfer for the DCD is shown in Fig.~\ref{f:TwoNExcite}.  The ground state (not shown) is only weakly populated due to Q-value mismatching, and thus is not further discussed.  Angular distributions for the closely-spaced $^{14}$C bound states could not be extracted due to the lack of $^4$He-$\gamma$ coincidences.  Hence, only the angular distribution for the $2^+_2$ 8.32 MeV state of $^{14}$C could be extracted. 

The optical potentials used in the transfer calculations are shown in table \ref{t:OM-Parameters}.  The elastic scattering parameters from \ref{elas} were used as the entrance channel and the $^5$He+$^{13}$C interaction potentials.  For the exit channel and the core-core potential, $\alpha$+$^{12}$C data at 28.2 MeV \cite{Mik61} was fit using {\sc sfresco}.  For the n+$^{12}$C binding potential, we took a  standard radius and diffuseness, and adjusted the depth to reproduce the correct binding energy.  A Gaussian potential was used for the $\alpha$+n binding potential \cite{Zhu93} and  finally the $^5$He+n binding potential was taken from Keeley {\it et al.} \cite{Kee08}.

\begin{figure}
\centering
\includegraphics[width=\columnwidth]{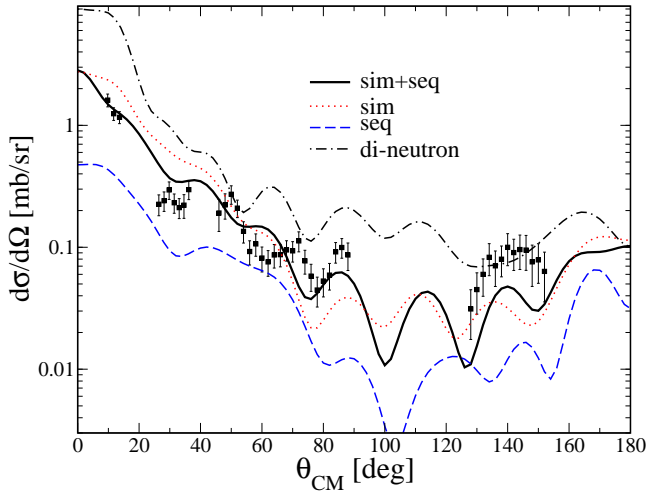}
\caption{\small (color online) Comparison of the angular distribution of the $^{12}$C($^6$He,$^4$He)$^{14}$C 2$^+$ $\mathrm{E}_{\mathrm{x}}=8.32$ MeV state at $\mathrm{E}_{\mathrm{lab}}$=30 MeV with the current model.  The dotted red line is the simultaneous two-neutron transfer accounting for the three-body nature of $^6$He, the dashed blue line is the sequential two-step transfer accounting for the structure of $^{13}$C and $^{14}$C and the solid black line is the coherent sum of the simultaneous and sequential transfer.  For comparison, the dot-dashed black line is the simple di-neutron model.  See text for details.  }
\label{f:transfer}
\end{figure}

The $2^+_2$ 8.32 MeV transfer cross-section is shown in Fig.~\ref{f:transfer}.  Note that only statistical errors are included for the data. The reaction calculations agree fairly well with the data, considering no normalization factor is applied to the transfer (no unhappiness factor).   The solid black line corresponds to the coherent sum of the sequential and simultaneous transfer, the dotted red line is the simultaneous transfer alone and the dashed blue line is the results of including the sequential transfer contributions only.

We performed various sensitivity tests to assess the robustness of our results. We found that taking the intermediate levels as degenerate introduces a reduction of the cross section by no more than $\approx 5\%$. However, the transfer cross section is by far most sensitive to the choice of the optical potentials. Taking a different set of parameters for an optical potential describing the $^6$He+$^{12}$C elastic scattering, resulted in significant effects on the magnitude (up to $50$\%) as well as on the shape of the angular distribution.  Effects of similar magnitude were seen for changes in the exit optical potential and in the core-core interaction. Less significant were the effects of the geometry used for the binding potentials of $^{12}$C+n/$^{13}$C+n.  In all cases the simultaneous two-neutron transfer remained the dominant component.  However, the interplay between the simultaneous and sequential transfer provides a better overall agreement with the data.

\begin{figure}
\centering
\includegraphics[width=\columnwidth]{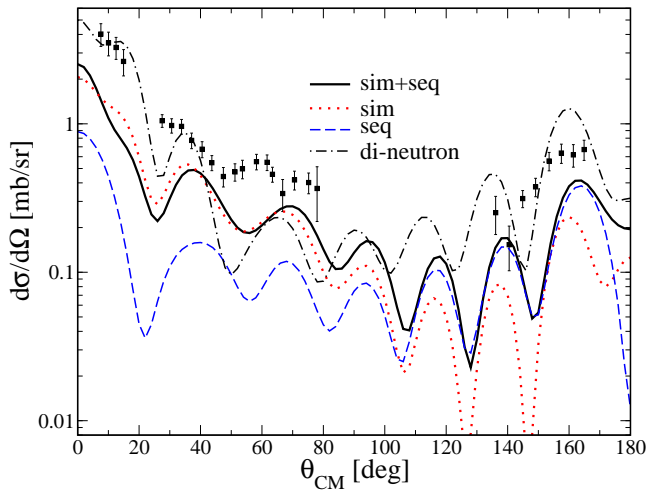}
\caption{\small (color online) Comparison of the angular distribution of the $^{12}$C($^6$He,$^4$He)$^{14}$C 2$^+$ $\mathrm{E}_{\mathrm{x}}=8.32$ MeV state at $\mathrm{E}_{\mathrm{lab}}$=18 MeV with the current model.  Refer to Fig.~\ref{f:transfer} for the description of the lines and the text for details.  } 
 \label{f:MilMod}
\end{figure}

For completeness, we examine the same reaction at 18 MeV \cite{Mil04}.  The entrance channel ($^6$He+$^{12}$C), exit channel ($^4$He+$^{14}$C) and sequential channel ($^5$He+$^{13}$C) were adjusted to those used by Milin {\it et al.} \cite{Obe68}. All structure information introduced in this calculation was kept the same as that used for the analysis of the 30 MeV reaction, for a meaningful comparison. As can be seen, the angular distributions presented in Fig.~\ref{f:MilMod} show a systematic underestimation of the experimental cross section.

Comparison with the pure di-neutron model was made for both the 30 MeV data and the 18 MeV data.  We use the same binding potentials in Milin {\it et al.} \cite{Mil04}, for $^{12}$C+2n and $^4$He+2n, keeping the optical potentials unchanged.  In $^6$He, we assumed the two neutrons were transfered from a relative 2S state.  The results obtained with the di-neutron model are shown in Figures \ref{f:transfer} and \ref{f:MilMod} by the dot-dashed line.  
Applying a renormalization of $\approx 0.3$ to the 30 MeV di-neutron cross section (not shown in the Fig.~\ref{f:transfer}) produces a distribution that resembles the respective data. The di-neutron model  appears to better describe the 18 MeV data, particularly at forward and backward angles without any normalization. Also, a coupled channel analysis  of the 18 MeV data \cite{Boz08}  including elastic, inelastic and transfer channels, again based on the simple di-neutron model, produces cross sections in agreement with the data. However, the agreement of the di-neutron model with the 18 MeV data is deceptive.  The fact that our simultaneous transfer predictions (dotted lines in Fig.~\ref{f:transfer} and \ref{f:MilMod}) are far from the di-neutron predictions (dot-dashed lines) stress the need for the inclusion of the correct three-body description of the projectile since it introduces important dynamics in the process.

While it is reassuring that the best reaction model is able to describe our transfer data at 30 MeV,  the factor of 2 mismatch between our best reaction model and the data at 18 MeV, shown  in Fig.~\ref{f:MilMod} calls for further investigation. Ideally, as a first step, one should perform a more thorough study of the optical potential parameter sensitivities at this energy. Next, one should study the effect of inelastic and continuum channels in the reaction mechanism. Such work is, however, well beyond the scope of the present paper.
\section{Conclusion}
\label{concl}
In summary, the elastic scattering of $^6$He on $^{12}$C at 30 MeV was measured and an angular distribution extracted.  
Inelastic scattering data was also extracted and analyzed including the quadrupole deformation of $^{12}$C. Qualitative agreement to the data was found, when accounting for full coupling of the ground state and first excited state of $^{12}$C.  Data for the two-neutron transfer angular distribution to the $2^+_2$ 8.32 MeV state in $^{14}$C was observed and analyzed, including the simultaneous and the sequential contributions.
A realistic three-body structure for $^6$He  was taken into account and state-of-the-art shell model predictions were used for the structure of the carbon isotopes.  

Overall, the reaction model describes the new data without invoking a normalization constant.  It was observed that the simultaneous transfer is the dominant reaction mechanism, yet the strong interplay between the intermediate states of $^{13}$C have non-negligible contributions to the final results.  This implies that adding the two-step process is required in order to better account for the overall reaction mechanism.  Discrepancies observed when adopting the current model to data at 18 MeV show that further theoretical work is required for a general and reliable approach to ($^6$He,$^4$He) reactions.  Even with the simultaneous experimental measurement of the elastic scattering, we find that the largest source of uncertainties in the calculation lies in the determination of the optical model potentials. Our sensitivity studies suggest that measurements of the elastic scattering all the way to $170^\circ$ could reduce these uncertainties.  This, however, presents some serious experimental challenges.

Concerning the reaction theory for the two-neutron transfer, this calls for further developments.  Future work should include the study of the effects of the continuum (e.g. the $2^+$ $^6$He resonance, the $^5$He states, etc)  in the reaction dynamics.  The strong sensitivity to the choice of optical potential parameters may be greatly reduced, if these are derived microscopically.
Finally, comparisons to multiple final bound states of $^{14}$C, such as the first $0^+$ and $2^+$ states, would allow a more thorough understanding of the spin dependencies of one-step and two-step processes.  

\begin{acknowledgments}

We would like to thank the staff of the ISAC-II facility for their efforts in delivering the $^6$He beam.  This work was partially supported by the US Department of Energy through Grant/Contract {Nos.} DE-FG03-93ER40789 (Colorado School of Mines) and DE-FG52-08NA28552, NSF grant PHY-1068217, and the Natural Sciences and Engineering Research Council of Canada.  TRIUMF is funded via a contribution agreement with the National Research Council of Canada.
\end{acknowledgments}

%

\end{document}